\documentclass{article}
\usepackage{amssymb}
\usepackage{amsmath}
\usepackage{amsthm} 
\usepackage{multirow}
\usepackage[english]{babel}
\usepackage{microtype}
\usepackage{siunitx}
\usepackage{algorithm}
\usepackage[noend]{algpseudocode}

\usepackage{graphicx}
\usepackage{array}

\usepackage[labelfont=bf]{caption}
\usepackage{float}
\usepackage{subcaption}
\usepackage[dvipsnames]{xcolor}
\usepackage[colorlinks]{hyperref}

\hypersetup{
colorlinks=true,
linktoc=all,
linkcolor=black,
citecolor=black,
urlcolor=black,
filecolor=black
}

\newcommand{\R}{\mathbb{R}}
\makeatletter
\newcommand\footnoteref[1]{\protected@xdef\@thefnmark{\ref{#1}}\@footnotemark}
\makeatother

\usepackage[margin=0.25in,headheight = 0.0in, footskip=0.2in]{geometry}  

\newtheorem{theorem}{Theorem}

\newtheorem{definition}{Definition}

\newtheorem{pro}[theorem]{Proposition}

\usepackage[shortlabels]{enumitem}

\title{\textbf{\huge Filaments and voids in planar central configurations}}

\author{Manuel R. Izquierdo\footnote{manuel.rodriguez@uib.es}\\
\normalsize
\textsf{IMCCE, Observatoire de Paris, 77, avenue Denfert-Rochereau, 75014 Paris, France} \\ \normalsize \textsf{Departament de Física, Universitat de les Illes Balears, Palma de Mallorca, Baleares, E-07122, Spain} \\ \normalsize \textsf{Institut Aplicacions Computationals (IAC3), Palma de Mallorca, Baleares E-07122, Spain}
}

\begin{document}
\maketitle

\begin{abstract}We have numerically computed planar central configurations of $n=1000$ bodies of equal masses. A classification of central configurations is proposed based on the numerical value of the complexity, $\mathcal{C}$. The main result of our work is the discovery of filaments and voids in planar central configurations with random complexity values. Suggestions are given for future work in the context of central configurations with random complexity values.
\end{abstract}

\tableofcontents


\section{Introduction}

\emph{Central configurations} (CC's) are a special class of configurations that give rise to the only known ``explicit'' solutions of the $n$-body problem. Regardless of its importance, only \emph{planar CC's} (PCC's) of a low number of bodies have been computed \textbf{\cite{moe89, fer02, mz19, dzd20}}, as well as \emph{spatial CC's} of $n=5, 6$ \textbf{\cite{mz20}} and $n=500$ bodies of \emph{equal masses} (SCCe's) \textbf{\cite{bgs03}}. In this work, we have numerically computed PCC's of $n=1000$ bodies of \emph{equal masses} (PCCe's). Classifying these CC's by the numerical value of the \emph{complexity} function, $\mathcal{C}$, we have observed filaments and voids in PCCe's of $n=1000$ with different complexity values.

In section \textbf{\ref{n_body}}, we briefly review the $n$-body problem and the main features of CC's. Later, in section \textbf{\ref{n = 1000}}, we present the computational scheme, as well as several PCCe's of $ n = 1000 $ with different complexity values. Finally, in section \textbf{\ref{discussion}} we expose the main unknowns derived from this work.

\section{The \emph{n}-body problem}\label{n_body}

The $n$-body problem aims to determine the possible motions of $n$ point particles of masses $m_{1}, \ldots, m_{n}$ that follow Newton's inverse square law, that is, the characterization of the dynamics of a system influenced by the gravitational interaction in the \emph{classical} regime. 

More precisely, if $q_{1}, \ldots, q_{n}$ represent the \emph{positions} at a given time $t$ of $n$ point particles with respective masses $m_{1}, \ldots, m_{n}$, their motion will be determined by the following second-order nonlinear differential equation

\begin{equation}\label{newton}
    \frac{d^{2}q_{i}}{dt^{2}} + \gamma_{i} \equiv \Ddot{q_{i}} + \gamma_{i} = 0  
\end{equation}
where

\begin{equation*}\label{gamma}
    \gamma_{i} = \sum_{k \neq i}m_{k}S_{ik}(q_{i} - q_{k}), ~~~~~
    S_{ik}=S_{ki}={\lVert q_{i} - q_{k} \lVert}^{-3}.
\end{equation*}

These bodies will lie in a $d$-dimensional Euclidean space, $q_{i} \in \R^{d}$. The \emph{energy}, $h$, is the difference of the \emph{kinetic energy}, $T$, and the \emph{force function} (opposite sign of the gravitational potential energy), $U$. For $G = 1$, it can be written as

\begin{equation*}
    h = T - U = \frac{1}{2}\sum_{i=1}^{n}m_{i}\dot{q_{i}}^{2} - \sum_{i<j} \frac{m_{i}m_{j}}{\lVert q_{i} - q_{j} \lVert}.
\end{equation*}

If we restrict ourselves to $d=3$ we find that $q, \dot{q} \in \R^{3n}$. Then, the $n$-body problem will be a $6n$ system of first-order equations. A complete solution would require $6n-1$ time-independent integrals and a time-dependent integral.

\subsection{Central configurations (CC's)}

In the context of the $n$-body problem, there are some privileged configurations called \emph{central configurations} (CC's). They are configurations of a particular type of solutions that are obtained if the point particles satisfy certain initial conditions. We call \emph{homographic solutions} those, such that the configuration formed by $n$-bodies at the instant $t$ remains similar to itself as time passes, up to dilations, rotations and translations.

\begin{definition}\label{CC}
A configuration is central if there exists a vector $\gamma_{O}$, a point $q_{O}$, and a $\lambda \in \R$ such that for all $i$, $\gamma_{i} - \gamma_{O}= \lambda (q_{i} - q_{O})$.
\end{definition}

In an equivalent way, we could say that it is a particular configuration where the position and acceleration vectors are proportional with the same constant of proportionality. This constant of proportionality is $\lambda$ which can be seen as a \emph{Lagrange multiplier}. Their main property is that they are the configurations that collapse \emph{homothetically} at their center of mass when released without initial velocity. For more details about CC's, see \textbf{\cite{alb03, moe90, saa05}}.

\subsection{Complexity}\label{complexity}

Motivated by the fact that the number of non-equivalent CC's increases extremely quickly as a function of $n$ \textbf{\cite{alb15}}, we propose a quantity that allows us to sort them in some way. Starting from the hypothesis that there are no non-equivalent CCe's with the same complexity values, we find an invariant quantity intrinsically related to the \emph{configurational measure}, $U \sqrt{I}$. Since $U$ is a positively homogeneous function\footnote{We say that a function $U: \R ^{m} \to \R$ is \emph{positively homogeneous of degree $\alpha$} if $U(tq)=t^{\alpha} U(q), \forall t \in \R^{+}, \forall q \in \R^{m}$. } of degree $-1$, and the \emph{moment of inertia}, $ I = \sum m_{i} q_{i}^{2}$, which describes the size of the system, is an homogeneous function of degree $2$; $U \sqrt{I}$ is an homogeneous function of degree $0$ whose value only depends on the shape (not on the size). In order to make it invariant of the scaling transformations of configuration space and masses, we will define it as:
\begin{equation}\label{complexity_equation}
    \mathcal{C} = \frac{U\sqrt{I}}{(\sum_{i} m_{i})^{5/2}}
\end{equation}

\section{Central configurations of \emph{n} = 1000 bodies of equal masses}\label{n = 1000}

\subsection{Computational scheme} 

\begin{figure}[H]	
\centering
\includegraphics[width=13 cm, height=10 cm]{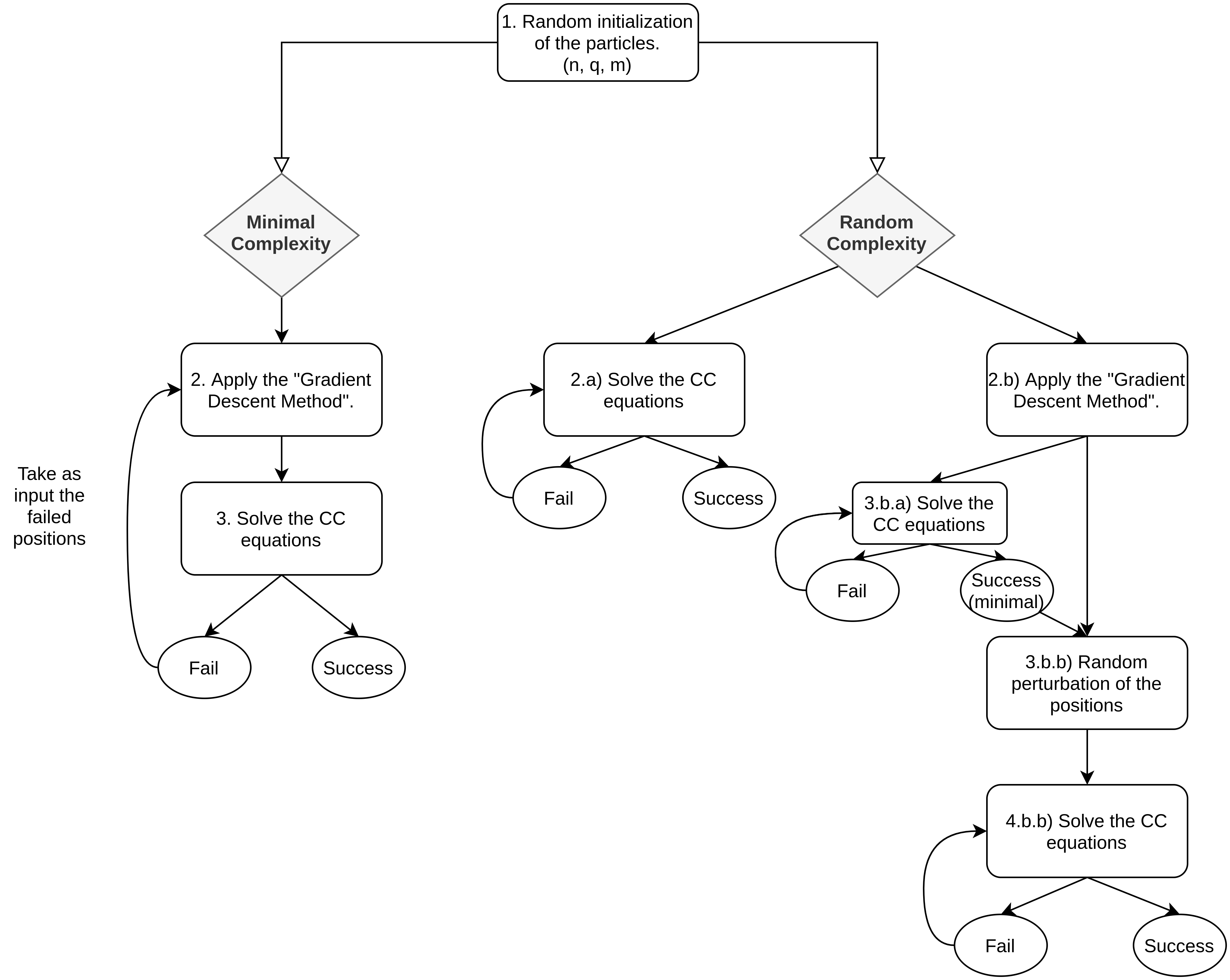}
\caption{Algorithm to compute PCCe's of $n$ bodies with minimal or random complexity values.}\label{computational_scheme}
\end{figure}

We present several PCCe's of $n=1000$ with $m=0.001$. In Fig. \textbf{\ref{computational_scheme}} the computational scheme that we follow is presented. First, we randomly initialize the particles, i.e. we assign them random positions. Then, we have to choose the level of \emph{complexity}. In our algorithm, we can choose if we want a CC corresponding to minimal or random complexity value. By \emph{random} we mean a CC that does not have a very low or very high complexity value.

Taking into account that CC's are critical points of the restricted Newtonian potential, $U$, with $I =1$, they are described by

\begin{equation*}
\frac{\partial U}{\partial q_{i}} + \frac{\lambda}{2} \frac{\partial I}{\partial q_{i}} = 0.
\end{equation*}

Without loosing any kind of generality, we can set $\lambda=1$ and minimize the quantity $U + \frac{1}{2} I$. For this task we have used the \emph{gradient descent} method. It follows as:

\textbf{Inputs:}
\begin{itemize}[noitemsep,topsep=0pt]
    \item \texttt{$t_{k}$}: Step size.
    \item \texttt{$\epsilon$}: Stopping condition.
\end{itemize}

\begin{algorithm}
\caption{Gradient Descent}
\begin{algorithmic}[1]
\State Guess $\mathbf{x}^{(0)}$, set $k = 0$
\While{$||\nabla f ( \mathbf{x}^{(k)}) || \geq \epsilon$}
    \State $ \mathbf{x}^{(k + 1)} = \mathbf{x}^{k} - t_{k} \nabla f (\mathbf{x}^{(k)})$
    \State $ k = k + 1$
\EndWhile
\State \Return $\mathbf{x}^{(k)}$
\end{algorithmic}
\end{algorithm}

The next step is to solve the set of $2n$ second order nonlinear differential equations which sets the CC positions. The computational challenge is caused by the scaling of the total terms as $n^{2}$. These equations greatly simplify if we assume that the total mass of the system $M = \sum m_{i}$ is non-zero.

\begin{pro}
Let $M=\sum m_{i} \neq 0$. By defining the center of mass as $q_{G} = \frac{1}{M} \sum m_{i} q_{i}$, a configuration is central if and only if there exists a $\lambda \in \R$ such that $\gamma_{i} = \lambda (q_{i} - q_{G})$.
\end{pro}

By setting $\lambda=1$ and $q_{G}=0$ we only have to solve the set of eqs. $\gamma_{i} - q_{i} = 0$. This purpose has been achieved through the \textsf{MINPACK} subroutine \texttt{hybrd}, which allows us to solve a set of $N$ nonlinear differential equations with $N$ variables by using a modification of Powell's hybrid method \textbf{\cite{pow1, pow2}}. Documentation can be found at \textbf{\cite{doc_minpack}}. If the interested reader wishes to replicate the numerical results obtained in this work, she should focus on the optimization of \texttt{FCN}, which is the user-supplied subroutine which calculates the functions and choosing an initial estimate of the solution vector (array of length $N$), \texttt{X}, which is close to the \texttt{FINAL APPROXIMATE SOLUTION}. Also, the relative error between two consecutive iterates, \texttt{XTOL}, can be tweaked if the convergence is too slow. We have checked the accuracy of our simulations by requiring:
\begin{itemize}[noitemsep,topsep=0pt]
    \item $|q_{G}| \leq [10^{-8}, 10^{-8}]$
    \item For all $i$, we require that $| \gamma_{i} - \lambda (q_{i} - q_{G} ) | \leq 10^{-8}$.
\end{itemize}

\subsection{CCe's with minimal complexity}

Although CC's are well known within the $n$-body problem, only lists of PCCe's up to $n\leq 12$ have been computed \cite{dzd20}. Previously, Ferrario \textbf{\cite{fer02}} presented a list of PCCe's of $n \in [3, 10]$. To test our numerical code, we have found all the figures of \textbf{\cite{fer02}}, as well as the three missing PCCe's of $n=10$ found by Doicu et al. \textbf{\cite{dzd20}}. These authors have designed different computational algorithms to be able to find the exact number of PCCe's for a given $n$. Our motivation is different. We wanted to compute PCCe's for a larger number of bodies. Here, we present PCCe's of $n = 1000$ with minimal complexity, a slightly greater value of complexity and an extremely high value of complexity. The lower bound of the numerical value of complexity are unknown, so we cannot be sure that these CCe's are an absolute minimum. We hypothesize that the PCCe with the highest value of complexity corresponds to the collinear case, i.e. when all the bodies are perfectly aligned. This assumption is motivated by Lindstrom's result in 1996 \textbf{\cite{lindstrom1, lindstrom2}}, in which he showed that in the limit $n \to \infty$ the value of $\mathcal{C}$ for configurations of minimal complexity is only unbounded for the collinear case, which scales as $\sim \log(n)$. Contrary to intuition, even in the case of equal masses, the bodies are not uniformly distributed, instead the density is greater in the center than at the extremes \textbf{\cite{lindstrom2}}. The density function, $f(x)$, is defined by

\begin{equation*}
  f_{\text{collinear}}(x)=\left\{
  \begin{array}{@{}ll@{}}
    \frac{3\sqrt{5}}{20}(1-\frac{x^{2}}{5}): & |x| \leq \sqrt{5} \\
    0: & |x| > \sqrt{5}
  \end{array}\right.
\end{equation*} 

\begin{figure}[H]	
\centering
\includegraphics[width=14 cm, height=2.5 cm]{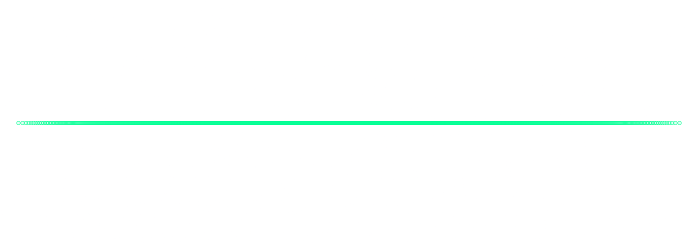}
\caption{The collinear CC of $n=1000$ bodies of equal masses. It has extremely high complexity, $\mathcal{C} = 1.78997168$.}\label{collinear}
\end{figure}

 In Fig. \textbf{\ref{collinear}} we have the collinear PCCe of $n=1000$, which is expected to have the highest value of complexity.
 
\begin{figure}[H]	
\centering
\includegraphics[width=9 cm,trim=4 4 4 4,clip]{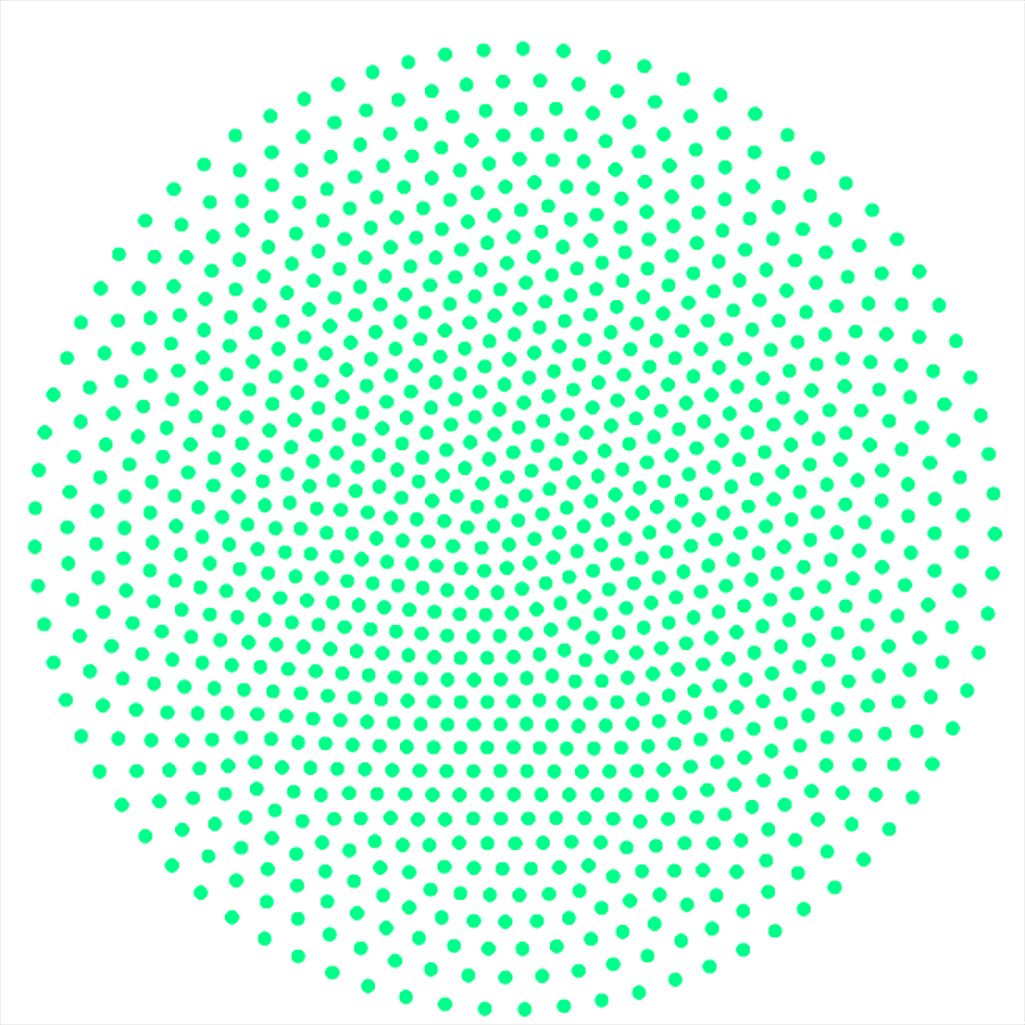}
\caption{Planar CC of $n=1000$ bodies of equal masses with minimal complexity, $\mathcal{C} = 0.57280571$.}\label{planar_low}
\end{figure}

The PCCe with the lowest value of complexity, that we found, can be seen in Fig. \textbf{\ref{planar_low}}. The numerical value obtained for the continuous limit, i.e. when $n \to \infty $, in theorem $4.2$ \textbf{\cite{lindstrom1}} is $\mathcal{C} \approx 0.59608$, which agrees quite well with our minimal PCCe. The density of PCCe's with minimal complexity is greater in the inner center than in the outer layers \textbf{\cite{lindstrom1}}. Therefore, they are not homogeneous, and also it is not symmetric. The density function, $f(r)$, has the following dependency of the polar radius, $r$,

\begin{equation*}
  f_{\text{planar}}(r)=\left\{
  \begin{array}{@{}ll@{}}
    \frac{3}{5\pi}\sqrt{1-\frac{2}{5}r^{2}}: & r \leq \sqrt{5/2} \\
    0: & r > \sqrt{5/2}
  \end{array}\right.
\end{equation*}

According to our numerical experiments and previous tests in the literature \textbf{\cite{fer02, mz19, dzd20}}, collinear CCe's are expected to have the highest complexity value. However, we highlight that this result has not been proven theoretically or numerically. If you compare the complexity values of Figs. \textbf{\ref{collinear}} and \textbf{\ref{planar_low}}, the former is approximately $ 3.13$ times greater than the latter\footnote{For PCCe's of $ n = 10 $, it is assumed that the PCCe of less complexity has a value of $\mathcal{C} \approx $ \num{3.66 e-6}, while the collinear has a value of $ \mathcal{C} \approx$ \num{5.49 e-6}. Therefore, the difference in this case is approximately $ 1.50$ times. Our numerical results agree with an 8-digit precision with those reported in \textbf{\cite{fer02}}.}.

\subsection{CCe's with random complexity values}

\begin{figure}[H]	
\centering
\includegraphics[width=8 cm]{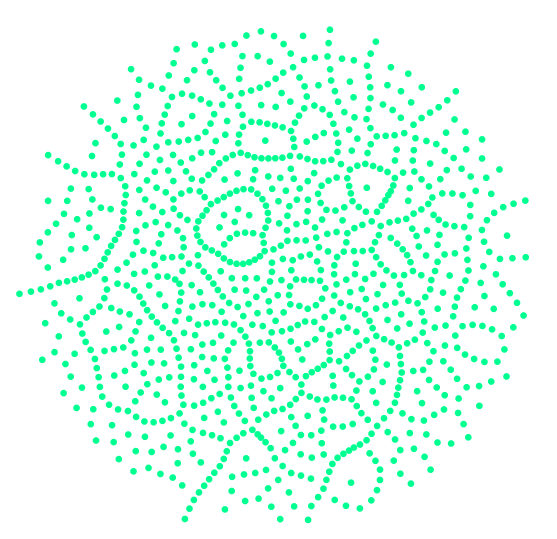}
\caption{Planar CC of $n=1000$ bodies of equal masses,  $\mathcal{C} = 0.57560474$.}\label{1000_1}
\end{figure}

\begin{figure}[H]	
\centering
\includegraphics[width=10.5 cm]{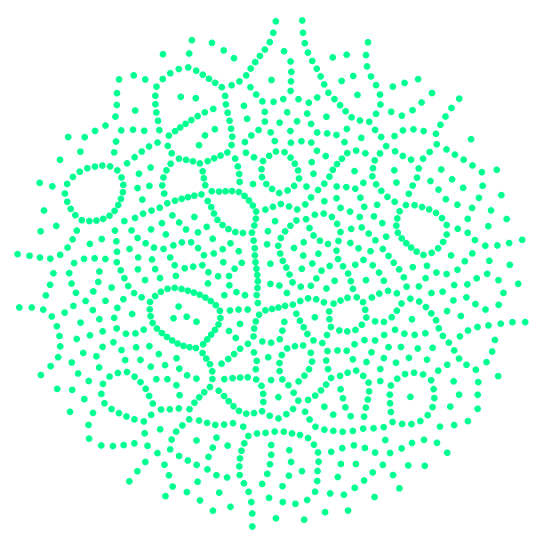}
\caption{Planar CC of $n=1000$ bodies of equal masses, $\mathcal{C} = 0.57640016$.}\label{1000_4}
\end{figure}

In Figs. \textbf{\ref{1000_1}} - \textbf{\ref{1000_4}} we show some PCCe's of $n=1000$ ordered in increasing level of complexity. We can easily observe that as CCe's get more complex, filaments and voids are formed. This unexpected result is one of the most interesting aspects of our work. We don't know exactly why are they formed. In fact, Fig. \textbf{\ref{1000_4}} is the PCCe of $n=1000$ with the highest complexity value that we found (apart from the collinear CCe of $n=1000$, in Fig. \textbf{\ref{collinear}}). It has a complexity value only around $1.006$ times greater than the PCCe of Fig. \textbf{\ref{planar_low}}. Unfortunately, we are closer to the absolute minimum than to the maximum (of Fig. \textbf{\ref{collinear}}).

\begin{figure}[H]	
\centering
\includegraphics[width=8.4 cm]{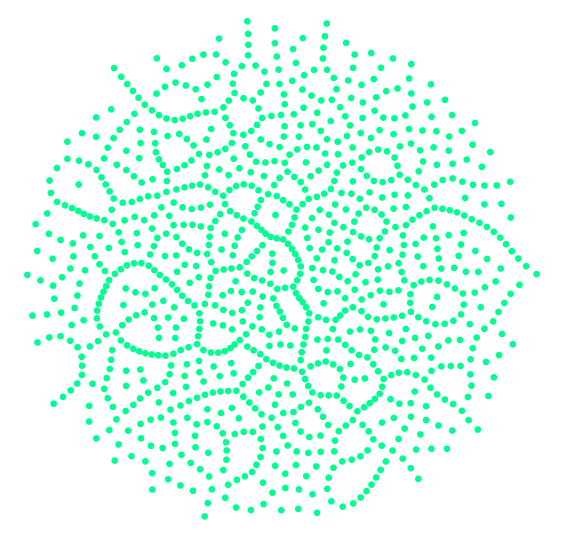}
\caption{Planar CC of $n=1000$ bodies of equal masses, $\mathcal{C} = 0.57574481$.}\label{1000_2}
\end{figure}

\begin{figure}[H]	
\centering
\includegraphics[width=10.5 cm]{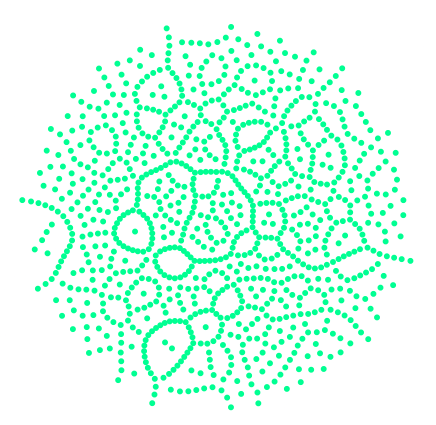}
\caption{Planar CC of $n=1000$ bodies of equal masses, $\mathcal{C} = 0.57599356$.}\label{1000_3}
\end{figure} 

\section{Discussion}\label{discussion}

By numerically computing PCCe's of $n=1000$ we have observed the formation of filaments and voids in configurations with random complexity values. Many questions have arisen from this discovery. The main unknowns derived from this work are:

\begin{enumerate}
    \item Why do filaments and voids form in PCCe's with random complexity values?
    \item Are there non-equivalent CCe's with the same numerical value of complexity?
    \item What is the numerical value of the CCe's with the highest complexity for a given $n$? Is it a collinear configuration?
    \item In Fig. \textbf{\ref{planar_low}} we have computed a PCCe of $n=1000$ with a very low value of complexity. Also, we hypothesize that Fig. \textbf{\ref{collinear}} corresponds to the PCCe of $n=1000$ with the highest complexity value. The second most complex PCCe of $n=1000$ that we found is in Fig. \textbf{\ref{1000_4}}. It is obvious that there should be many CC's between Fig. \textbf{\ref{1000_4}} and Fig. \textbf{\ref{collinear}}, how do they look like? 
\end{enumerate}

\section*{Acknowledgments}
My sincere thanks to Alain Albouy for showing me the beauty of central configurations. As well as Jacques Féjoz for computational advise. Finally, I really appreciate the multiple discussions with Julian Barbour, who has definitely given my work a completely different perspective.

\section*{Funding}
This research was funded by the Department of Mathematics of Université Paris Dauphine-PSL. MRI also acknowledges financial support from the Laboratoire d’Excellence UnivEarthS.

\section*{Abbreviations}
The following abbreviations are used in this manuscript:\\
\noindent 
\begin{tabular}{@{}ll}
CC & Central configuration\\
CCe & CC of bodies of equal masses ($m=0.001$) \\
CC's & Central configurations\\
PCC & Planar CC\\
PCCe & Planar CC of bodies of equal masses ($m=0.001$)\\
SCC & Spatial CC\\
SCCe & Spatial CC of bodies of equal masses ($m=0.001$)
\end{tabular}

\bibliographystyle{apalike}

\end{document}